\documentclass[aps,prc,twocolumn,showpacs,twoside]{revtex4}
\usepackage{bm,graphicx}

\begin{document}

\title{Planar versus non-planar $\bm{\bar NN}$ annihilation into mesons in the light of $\bm{\bar qq}$ operators and the $\bm{1/N_c}$ expansion}

\author{B.~El-Bennich}
\affiliation{Laboratoire de Physique Nucl\'eaire et Hautes \'Energies, Groupe Th\'eorie \\ 4, place Jussieu, Universit\'e Pierre et Marie Curie, 75252 Paris, France}
\date{\today}

\newcommand{\s}{^3S_1}
\newcommand{\p}{^3P_0}
\newcommand{\pvec}{{\mathbf p}}
\newcommand{\etal}{{\em et al.\/}}


\begin{abstract}
It is argued that in the antiproton-proton annihilation into two mesons $\bar pp\rightarrow m_1 m_2$, the origin of different
restrictive angular momentum selection rules commonly obtained for planar annihilation diagrams $A2$ and for non-planar 
rearrangement diagrams $R2$ lies in the omission of momentum transfer between an annihilated antiquark-quark pair and 
a remaining quark or antiquark. If it is included, there is no reason for dismissing one type of diagram in favor of another
one. Some considerations in the large $N_c$ limit of QCD equally shed light on the planar and non-planar contributions 
to the total $\bar NN\rightarrow m_1 m_2$ annihilation amplitude. 
\pacs{12.39.Jh, 13.75.Cs, 21.30.Fe, 25.43.+t}
\end{abstract}

\maketitle

\section{The Problem with $\bm{R2}$ versus $\bm{A2}$ Diagrams \label{sec1}}

\begin{figure*}[t]
\includegraphics[scale=0.71]{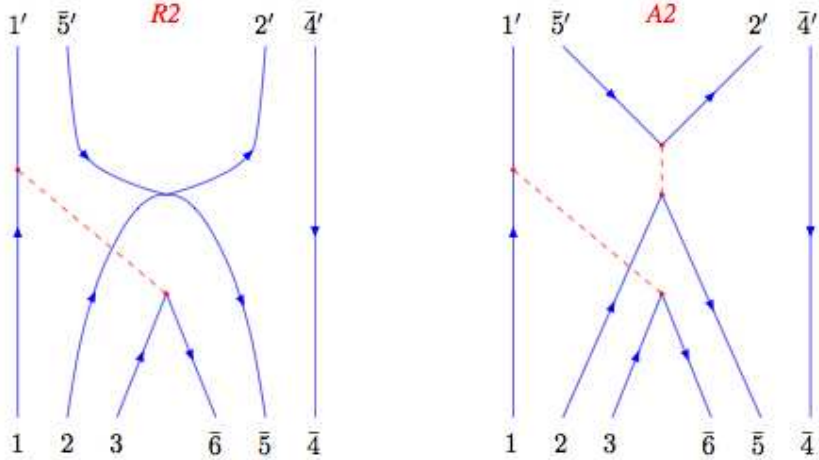}

\caption{Rearrangement diagram $R2$ (left) and annihilation diagram $A2$ (right). The numbers with bars denote antiquarks, those without 
bars the quarks. The dashed lines represent the exchange of an effective state with either vacuum $\p$ or gluon $\s$ quantum numbers 
and with momentum transfer $\delta (\pvec_{1'}-\pvec_1-\pvec_3-\pvec_6)$ for $R2$ and $\delta (\pvec_{1'}-\pvec_1-\pvec_3-\pvec_6)$
$\delta (\pvec_{2'}+\pvec_{5'}-\pvec_2-\pvec_5)$ for $A2$. }        
\label{diagrams}
\end{figure*}

In antiproton-proton annihilation described by quark-line diagrams, within the context of the constituent quark model, 
the common wisdom is that planar diagrams $A2$ and non-planar $R2$ diagrams do not contribute equally in a given reaction, 
say $\bar pp \rightarrow \pi\pi$ or $\bar pp \rightarrow \pi\rho$. As depicted in Fig.~\ref{diagrams}, both the $R2$ and 
$A2$ diagram can lead to the same final non-strange two-body configuration despite the different flavor-flux 
topology. Since the initial $\bar pp$ pair contains no strangeness, the annihilation $\bar pp \rightarrow \bar KK$ can 
proceed {\em directly\/} only via the $A2$ diagram. Indirectly, taking into account final-state interactions, the two-pion 
final state can couple to the $\bar KK$ state {\em via\/} the $R2$ topology.

Some authors \cite{hartmann} favor the $R2$ topology due to the following observation: for $A2$ one gets the same branching ratio 
(assuming SU(3) symmetry is unbroken) for $\bar pp \rightarrow \pi\rho$ and $\bar pp\rightarrow \bar KK^*$. The experimentally observed 
ratio $\mathrm{Br}(\bar KK^*)/\mathrm{Br}(\pi\rho)$ is small and favors the $R2$ topology which produces only $\pi\rho$ but 
not $\bar KK^*$ (where the authors ignore final-state interactions).  It is therefore concluded that among the two graphs $R2$ is the 
dominant one. However, in the review of nucleon-antinucleon annihilation by Dover \etal~\cite{dover}, it is noted that this argument 
ignores a strong mechanism of SU(3) breaking, namely the suppression of $\bar ss$ pairs \cite{vandermeulen}. It seems thus incorrect 
to attribute the suppression of strange modes $\bar KK$, $\bar KK^*$~{\em etc.\/} to the dominance of the $R2$ over the $A2$ topology. 
Moreover, the authors of the review \cite{dover} reverse the argument and reason in terms of selection rules to establish instead the 
preponderance of $A2$ over $R2$ diagrams in $\bar pp$ annihilation into two mesons.

The {\em pro}-$A2$ argument follows from a supposedly restrictive set of selection rules for the $R2$ diagrams, which does 
not allow for the experimentally observed annihilation of a $\bar pp$ pair into two pions (and similar restrictions also hold for 
the annihilation into two different mesons) with certain total angular momentum $J=l_{\bar pp}\pm 1=\ell_{\pi\pi}$. In more detail, if we 
concentrate on $\pi\pi$ final states, the $R2$ annihilation diagram with the specific rules of  Ref.~\cite{dover} permits 
only $S$-waves ($\ell_{\pi\pi}=0$) when the $\bar q_6 q_3$ pair annihilates into a vacuum $\p$ state, whereas $\bar q_6 q_3$  
annihilation into a $\s$ state with gluon quantum numbers restricts $\ell_{\pi\pi}$ to $S$- and $P$-waves. No final state with 
$\ell_{\pi\pi}=2$ or higher is allowed in either case. The planar $A2$ diagram, on the other hand, does not exhibit this restriction 
on the orbital angular momentum $\ell_{\pi\pi}$.

One argument speaks against these restrictions --- the specific rules applied by the authors of Ref.~\cite{dover} stem from
the absence of momentum transfer from the annihilated $\bar q_6 q_3$ vertex to any of the remaining quarks or antiquarks 
in Fig.~\ref{diagrams}, which we claim is not realistic. If, on the other hand, momentum transfer is allowed in the non-planar $R2$ diagram 
\cite{kloet,elbennich1,elbennich2,elbennich3}, the selection rules are modified and result into $R2$ transition operators
\begin{widetext}
\begin{eqnarray}
\hat T_{R2}(\p) &=& i\mathcal{N}\Big [ A_V \bm{\sigma}\!\cdot\!\mathbf{R'}\sinh(C\,\mathbf{R}\!\cdot\!\mathbf{R'}) +
 B_V \bm{\sigma}\!\cdot\!\mathbf{R}\cosh(C\,\mathbf{R}\!\cdot\!\mathbf{R'}) +  C_V (\bm{\sigma}\!\cdot\!\mathbf{\hat R'})\,R\cos\theta 
 \cosh(C\,\mathbf{R}\!\cdot\!\mathbf{R'})\Big ]\times \nonumber \\
 & & \times \; \exp \{A\mathbf{R'}^2+B\mathbf{R}^2+D \mathbf{R}^2\cos^2\theta \} .
\label{tot1} 
\end{eqnarray}
for $\bar qq$ annihilation into a $\p$ state. The $\s$ transition operator is split into a  {\em longitudinal} component
\begin{eqnarray}
\hat T_{R2}(\s^L) & = & i\mathcal{N}\Big [ A_L \bm{\sigma}\!\cdot\!\mathbf{R'}\sinh(C\,\mathbf{R}\!\cdot\!\mathbf{R'}) +
 B_L \bm{\sigma}\!\cdot\!\mathbf{R}\cosh(C\,\mathbf{R}\!\cdot\!\mathbf{R'}) +  C_L (\bm{\sigma}\!\cdot\!\mathbf{\hat R'})\,R\cos\theta 
 \cosh(C\,\mathbf{R}\!\cdot\!\mathbf{R'})\Big ]\times \nonumber \\
 & &\times \; \exp \{A\mathbf{R'}^2+B\mathbf{R}^2+D \mathbf{R}^2\cos^2\theta \} 
\label{tot2}
\end{eqnarray}
and a {\em transversal} component
\begin{eqnarray}
\hat T_{R2}(\s^T) & = & \mathcal{N}\Big [ A_T \bm{\sigma}\!\cdot\!\mathbf{R'}\cosh(C\,\mathbf{R}\!\cdot\!\mathbf{R'}) + 
 B_T \bm{\sigma}\!\cdot\!\mathbf{R}\sinh(C\,\mathbf{R}\!\cdot\!\mathbf{R'}) + C_T (\bm{\sigma}\!\cdot\!\mathbf{\hat R'})\,R\cos\theta 
 \sinh(C\,\mathbf{R}\!\cdot\!\mathbf{R'})\Big ]\times \nonumber \\
 & &\times \, \exp \{A\mathbf{R'}^2+B\mathbf{R}^2+D \mathbf{R}^2\cos^2\theta \}.
\label{tot3}
\end{eqnarray}
\end{widetext}
Here, $\mathbf{R'}=\mathbf{R}_{m_1}\!\!-\mathbf{R}_{m_2}$  and $\mathbf{R}=\mathbf{R}_{\bar p}-\mathbf{R}_p$ are the relative meson and 
antiproton-proton coordinates in the c.m. system, respectively. The c.m. angle $\theta$ is between the relative meson vector $\mathbf{R'}$ 
and antiproton-proton vector $\mathbf{R}$, $\bm{\sigma}$ are the usual Pauli matrices and $\mathcal{N}$ is an overall normalization. The 
coefficients $A_i, B_i, C_i$ with $i=V,L,T$ and $A, B, C$ and $D$ depend on size parameters $\alpha$ (proton), $\beta$ (pion) and the boost 
factor $\gamma=E_{\mathrm{cm}}/2m c^2$. They are detailed in Ref.~\cite{elbennich2}. Note that for $\gamma=1$ the coefficients correspond 
to those in Refs.~\cite{kloet,elbennich1}.  As was already discussed in Ref.~\cite{kloet} and its relativistic extension in Ref.~\cite{elbennich2}, 
sandwiching the transition operators between the two-pion and the $\bar NN$ wave functions, and taking into account the parity properties of the
particles, shows that $\hat T_{R2}(\p)$ and $\hat T_{R2}(\s^L) $ act in $\ell_{\pi\pi}$ even waves whereas $\hat T_{R2}(\s^T) $ contributes only to 
$\ell_{\pi\pi}$ odd waves. Hence, one obtains $\bar pp\rightarrow \pi\pi$ annihilation for $\ell_{\pi\pi}=0,1,2,3, ...$, that is in $S,P,D,F$,... 
waves, provided both the $\p$ and $\s$ mechanisms are taken into account. A similar analysis of the $A2$ diagrams yields exactly the same 
selections rules.

The conclusion is that selection rules do not discriminate between the two topologies, $R2$ and $A2$, which both describe the annihilation. 
So, within a constituent quark model in a non-perturbative regime, it is a more consistent approach to consider all annihilation amplitudes equally 
as they represent different aspects of QCD.

\section{$\bar NN$ Annihilation in the large $\bm{N_c}$ context}

We mention another approach that sheds some light on the question of dominance of either $R2$ or $A2$ topology, namely the treatment of 
$\bar pp$ annihilation in the large-$N_c$ limit of QCD. This has already been done previously~\cite{pirner} and in the following we
summarize some salient results that underpin our conclusion in Section~\ref{sec1}. Again, one wants to test, if not quantitatively then at least
qualitatively, whether a particular annihilation process is dominant in any reaction $\bar NN \longrightarrow m_1m_2$.

In his seminal paper \cite{thooft}, 't Hooft suggested that one can generalize QCD from three colors and an SU(3) gauge group to $N_c$ 
colors and an SU$(N_c)$ gauge group. It was shown that QCD simplifies in the limit of large $N_c$ and $g^2N_c$  ($g^2=4\pi\alpha_s$) 
fixed and that there exists a systematic expansion in powers of $1/N_c$. One crucial property of this limit is that at $N_c=\infty$ the meson 
and glue states are free, stable and non-interacting. Other important consequences are that mesons become {\em pure\/} $\bar qq$ 
states, since the sea quarks and antiquarks vanish for large $N_c$. Moreover, Zweig's rule is exact in this limit. These results 
are restricted to color-singlet glue states (glue balls) and mesons, where a clear distinction between planar and non-planar diagrams
is possible. This follows from the Euler index for each Feynman diagram which determines the $1/N_c$-counting. 

The extension to baryons was carried out by Witten, who treated the nucleon-nucleon interaction in large-$N_c$ QCD \cite{witten}.
The main result that emerged from his qualitative analysis is the $N_c$-order of elastic baryon-baryon and baryon-meson scattering
amplitudes. The former is of order $N_c$ and the latter of order $N_c^0$. Since it is shown in Ref.~\cite{witten} that all energy contributions 
to the baryon mass $M$ are of order $N_c$, the (non-relativistic) kinetic energy $Mv^2/2$ is of equal order $N_c$ as the baryon-baryon 
interaction energy. Hence, the scattering cross section for baryon-baryon scattering does not vanish in the large $N_c$ limit. 
On the other hand, the baryon-meson scattering amplitude, being of order $N_c^0$ according to Ref.~\cite{witten}, is negligible 
compared with the baryon mass and to leading order in $1/N_c$ the baryon propagates unperturbed by mesons. The meson mass is of 
order $N_c^0$; it follows that its kinetic energy is of the same order as the baryon-meson interaction energy, which is therefore large 
enough to influence the motion of the meson. For the dynamics of baryon-meson systems, this means the meson is scattered off the 
baryon but the baryon itself remains ``free" for $N_c\rightarrow \infty$.

\begin{figure*}[t]
\includegraphics[scale=0.8]{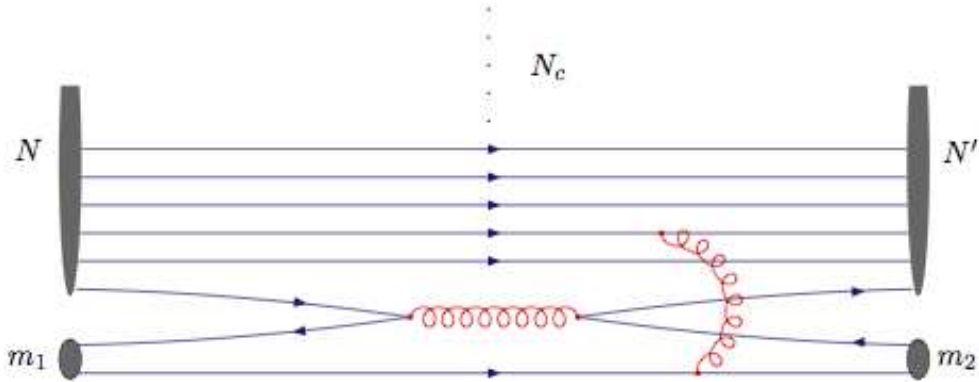}
\caption{Quark-line diagram of meson-nucleon scattering in the large $N_c$ limit. The quark-gluon vertices carry the coupling constant 
         $g/\sqrt{N_c}$. In principle, a gluon can be exchanged between the spectator quark in the meson and any other $N_c-1$ quarks in 
         the nucleon without altering the $N_c$-order of the scattering amplitude.} 
\label{fig2}
\end{figure*}

The aim of this section is to verify whether the above results can be applied to antinucleon-nucleon annihilation into two mesons. To begin with, 
it should be pointed out that for baryons the topological denominations {\em planar} and {\em non-planar} are a misnomer in the large $N_c$ 
context, as there exists no Euler index according to which the diagrams could be ordered. It is clear from Fig.~\ref{diagrams} that it is awkward 
to generalize $\bar NN$ annihilation into two mesons for $N_c$ quarks and $N_c$ antiquarks, for the simple reason that mesons remain 
$\bar qq$ states for large $N_c$. Nevertheless, assume a naive picture in which one first chooses out of the $N_c$ quarks and $N_c$ antiquarks
{\em one\/} quark and {\em one\/} antiquark which do not end up in the same meson. This gives a factor of $N_c^2$ and the color content of each
of the two final mesons is fixed. Next, a quark is rearranged with an antiquark and the two-meson final state is then formed with the first quark and 
antiquark choice as in Fig.~\ref{diagrams}. There is no additional color factor $N_c$ arising from this as the color-neutrality of the mesons demands
a unique choice in the rearrangement. A gluon exchange between the rearranged quark and antiquark does not alter the $N_c$ counting at
this stage (remember the colors in the two mesons were previously fixed), however it introduces an $1/N_c$ factor from the gluon couplings,
where each vertex carries a factor $g/\sqrt{N_c}$.

One can now proceed to annihilate {\em all\/} the remaining $\bar qq$ pairs. In principle, the $\bar qq$ pairs need not form a color singlet since 
a gluon from the annihilation vertex can be exchanged with any other still remaining (anti)quark. Therefore, each annihilation comes along with 
a factor $N_c-k$, where $k=2,3,4 ...$ is the number of quarks that have been annihilated, but each also involves a factor $1/N_c$ from the gluon 
vertices. In our case, we start with $k=2$ as two of the quarks (and antiquarks) are not annihilated and therefore end up in the final two mesons.
An equivalent derivation is to think of the $N_c-k$ quarks and $N_c-k$ antiquarks annihilated into vacuum states, each of which can be treated 
as a $\bar qq$ state. Since any permutation of pairing a quark with an antiquark is possible, this gives a factor $(N_c-k)!$ Using this result and the factor $N_c^2$ derived in the previous paragraph, one obtains an $N_c$-order for $\bar pp$ annihilation into $k$ mesons: 
$N_c^2 (N_c-k)! /N_c^{N_c-k}$. It follows from this that for $k=2$, the $R2$ diagram generalization to $N_c$ colors is of order
\begin{equation}
  N_c^2 \times \frac{1}{N_c^{N_c-2}} \times (N_c-2)!  \simeq  \frac{N_c!}{N_c^{N_c}}  \stackrel{N_c\rightarrow \infty}{\longrightarrow} e^{-N_c},
\label{order}
\end{equation}
where for large $N_c$ one approximates $N_c-2 \simeq N_c-4 \simeq N_c$. In the second step of Eq.~(\ref{order}), Stirling's approximation 
$n! \simeq \sqrt{2\pi}\, n^{n+\frac{1}{2}} \exp (-n)$ has been employed. 
Repeating this procedure for the $N_c$ generalization of the $A2$ diagram, one finds it is suppressed by a factor $1/N_c^2$ with respect
to the above discussed $R2$ case. The origin of this lies in the lack of quark-antiquark rearrangement, which contributes a non-vanishing 
$N_c$-factor without cancelling effect from the gluon vertices. Instead, here a $\bar qq$ pair has to be created from a gluon. One eventually 
arrives at an expression for the $1/N_c$-order very similar to Eq.~(\ref{order}) except that the first factor is $N_c$ rather than $N_c^2$, the 
denominator in Eq.~(\ref{order}) becomes $N_c^{N_c-1}$ since there is an additional factor due to the gluon coupling to the created 
$\bar qq$ pair, and for the same reason the last factor is  $(N_c-1)!$ instead of $(N_c-2)!$ Given the effect of the dominating pairwise 
$\bar qq$ annihilations in both types of diagrams, it can be concluded that either way the annihilation into two mesons is exponentially 
damped, which can also be deduced from the quark model calculations in Ref.~\cite{pirner}. We shall return to this shortly.

We allow ourselves a short digression at this point --- in the cross-channel reaction $m_1 N\rightarrow m_2 N$ one also encounters
a variety of processes contributing in the large $N_c$ limit to the scattering amplitude. An example of such scattering is illustrated in
Fig.~\ref{fig2}, where a nucleon-quark is annihilated with an antiquark of the initial meson. A $\bar qq$ pair is produced in the final state
and since color is transferred between the annihilation and the creation vertex, one gets a factor $N_c$ from choosing a quark in the
nucleon as well as a factor $1/N_c$ from the gluon couplings. The initial meson-quark is still contained in the final meson and acts as
a spectator. The scattering is therefore of order $N_c^0$. If instead of annihilation of a $\bar qq$ pair we resort to quark-rearrangement, 
that is a quark of the nucleon ends up in the final meson while the quark of the initial meson replaces this missing quark in the nucleon, 
we find the scattering amplitude to be of the same order $N_c^0$, as discussed in Ref.~\cite{witten}. This, in turn, differs from the 
$\bar NN\rightarrow m_1 m_2$ annihilation, where rearrangement was shown to yield an additional factor $N_c$ in the previous section.

Coming back to $\bar NN$ annihilation in the large $N_c$ limit, we recapitulate the results derived by Pirner~\cite{pirner}. In this approach,
overlap integrals of meson and $\bar NN$ wave functions generalized to $N_c$ quarks and $N_c$ antiquarks are worked out.
Including appropriately normalized color wave functions introduces a $N_c$ dependence from the color matrix elements. Furthermore, 
in order to obtain the annihilation cross section, one needs to calculate the $N_c$ dependent phase-space integral. With all these ingredients
the cross section for $\bar NN$ annihilation into $N_c-k$ mesons and quark-antiquark rearrangement behaves in the large $N_c$  limit as
\begin{eqnarray} 
   \sigma_{ \mathrm{ann.}} &\stackrel{N_c\rightarrow\infty}{\longrightarrow}& \exp (-4N_c\epsilon_0 r^2_m m) \times \nonumber \\
   & &\hspace*{-6mm} \times \exp  \{ [ 1/2- 3/2 \ln(3/2) ] N_c \}  (\eta N_c)^{2k}
\label{expdamp1}
\end{eqnarray}
where $m$ is the mass, $r_m$ the size, and $\epsilon_0$ the kinetic energy of the mesons (this holds only for same types of mesons like 
$\pi^\pm,\pi^0$ \emph{etc}.). The non-perturbative annihilation or creation probability is $\eta^2\simeq 0.25$. The cross section for 
$\bar NN$ annihilation into $N_c-k+l$ mesons, where $l$ is the number of created quarks, is for large $N_c$
\begin{eqnarray}
   \sigma_{ \mathrm{ann.}} &\stackrel{N_c\rightarrow\infty}{\longrightarrow}& \exp (-4N_c\epsilon_0 r^2_m m) \times \nonumber \\
   & & \hspace*{-6mm}\times \exp  \{ [ 1/2- 3/2 \ln(3/2) ] N_c \}  (\eta N_c)^{2k} \eta^{2l}\!.
\label{expdamp2}
\end{eqnarray}
Obviously,  Eqs.~(\ref{expdamp1}) and (\ref{expdamp2}) just differ by a factor $\eta^{2l}$. As $\eta < 1$, the pure annihilation diagram
without rearrangement is slightly more damped depending on the number $l$ of created quarks. This is in accordance with the remarks made 
above about the combinatorial factors in large $N_c$ generalizations of $R2$ and $A2$ diagrams. In the ``real world", where $N_c=3$, one can
at the most annihilate $k=3$ quarks (thus violating Zweig's rule) and create $l$ mesons. The more mesons are produced in the final state, the more
the rearrangement diagrams should dominate however slight this difference is.

\section{Conclusive Remarks}

The conclusions that can be drawn from the preceding two sections are coherent. Whether using a quark model calculation or a qualitative analysis
of the annihilation in the large $N_c$ limit of QCD, there is no evidence for dominance of either diagram, be it of the annihilation ($A2$)  or
rearrangement ($R2$) type. It should be mentioned that similar findings were reported in Refs.~\cite{elbennich1} and \cite{elbennich3}, where it 
was noted that the $R2$ diagrams suffice (although both annihilation mechanisms $\p$ and $\s$ are needed) to reproduce the LEAR data on 
$\bar pp \to \pi^-\pi^+$ differential cross sections and analyzing powers \cite{hasan}. This contradicts the statements of Dover \etal~\cite{dover}, 
who based their reasoning on SU(3) symmetry breaking and selection rules. Moreover, the present discussion can be generalized to annihilation 
of any two baryons or hyperons into two or more mesons.

\begin{acknowledgments}

This work was supported by a Marie Curie International Reintegration Grant under Contract No.~516228.  I would like to thank Wim Kloet, Beno\^{\i}t Loiseau 
and Boris Gelman for stimulating discussions.  Laboratoire de Physique Nucl\'eaire et Hautes  \'Energies is Unit\'e de Recherche des Universit\'es Paris 6 et 7 
associ\'ee au CNRS. 

\end{acknowledgments}

\end{document}